\documentclass[10pt,english,aip,apl,twocolumn]{revtex4-1}
\usepackage[T1]{fontenc}
\usepackage[latin9]{inputenc}
\usepackage{units}
\usepackage{amstext}
\usepackage{amssymb}
\usepackage{graphicx}
\usepackage{esint}

\makeatletter

\@ifundefined{textcolor}{}
{%
 \definecolor{BLACK}{gray}{0}
 \definecolor{WHITE}{gray}{1}
 \definecolor{RED}{rgb}{1,0,0}
 \definecolor{GREEN}{rgb}{0,1,0}
 \definecolor{BLUE}{rgb}{0,0,1}
 \definecolor{CYAN}{cmyk}{1,0,0,0}
 \definecolor{MAGENTA}{cmyk}{0,1,0,0}
 \definecolor{YELLOW}{cmyk}{0,0,1,0}
}

\makeatother

\usepackage{babel}
\begin{document}

\title{Spin Transport and Precession in Graphene measured by Nonlocal and
Three-Terminal Methods}

\author{André Dankert}

\email{andre.dankert@chalmers.se}

\affiliation{Department of Microtechnology and Nanoscience, Chalmers University
of Technology, SE-41296, Göteborg, Sweden}

\author{Mutta Venkata Kamalakar}

\affiliation{Department of Microtechnology and Nanoscience, Chalmers University
of Technology, SE-41296, Göteborg, Sweden}

\author{Johan Bergsten}

\affiliation{Department of Microtechnology and Nanoscience, Chalmers University
of Technology, SE-41296, Göteborg, Sweden}

\author{Saroj P. Dash}

\email{saroj.dash@chalmers.se}

\affiliation{Department of Microtechnology and Nanoscience, Chalmers University
of Technology, SE-41296, Göteborg, Sweden}
\begin{abstract}
We investigate the spin transport and precession in graphene by using
the Hanle effect in nonlocal and three-terminal measurement geometries.
Identical spin lifetimes, spin diffusion lengths and spin polarizations
are observed in graphene devices for both techniques over a wide range
of temperatures. The magnitude of the spin signals is well explained
by spin transport models. These observations rules out any signal
enhancements or additional scattering mechanisms at the interfaces
for both geometries. This validates the applicability of both
the measurement methods for graphene based spintronics devices and
their reliable extractions of spin parameters. 
\end{abstract}


\maketitle

\section*{}

The spin degree of freedom of electrons is considered as an alternative
state variable for processing information beyond the charge based
CMOS technology. Its potential lies in the possibilities for a new
generation of computers that can be non-volatile, faster, smaller,
and capable of simultaneous data storage and processing with a reduced
energy consumption \cite{Awschalom2007}. The strong interest in graphene
and silicon based spintronic devices stems from their potentially
long spin coherence lengths due to the absence of hyperfine interactions
and a weak spin-orbit coupling. Such materials could be employed in
the recently proposed concept of all spin logic using spins in ferromagnets
to store information and communicate between them using a spin current
\cite{Dery2007}. All spin logic is particularly powerful since it
combines various spin related phenomena such as spin injection, transport
and detection with magnetization dynamics. 

In order to achieve these goals various methods for electrical spin
injection and detection in metals \cite{Jedema2002}, semiconductors
\cite{Lou2007,VantErve2007,Dash2009} and graphene \cite{Tombros2007}
have been investigated. Primarily nonlocal (NL) and
three-terminal (3T) methods are used for an electrical detection of
the spin polarization \cite{Tombros2007,Dankert2013,Dankert2013a,Dash2009}.
The non-local geometry separates the current and voltage path to provide
information about pure spin transport parameters. However, nanofabrication
by electron beam lithography is necessary in order to achieve submicrometer
structures and channel lengths \cite{Tombros2007}. Although the NL
method has been widely used for spin transport measurements in more
conducting metals \cite{Jedema2002}, graphene \cite{Tombros2007}
and GaAs \cite{Salis2011,Lou2007}, it has been found to be challenging
for Si and Ge based devices due to the high resistive Schottky barriers.
Nevertheless, there are few reports studying NL signals in those materials
at low temperature \cite{Sasaki2011}, and more recently even at room
temperature \cite{Suzuki2011,Chang2013a,Sasaki2014}. Therefore, the
3T Hanle technique was prefered for measuring spin signals in semiconductor
materials, since it allows to study the creation and detection of
spin accumulations by a single magnetic tunnel contact up to room
temperature in a reproducible way \cite{Dash2009,Jain2012b,Dankert2013}.
\begin{figure}
\begin{centering}
\includegraphics[scale=1.25]{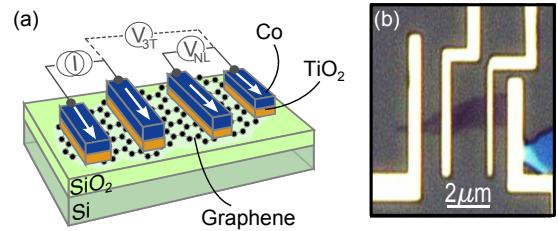} 
\end{centering}

\protect\caption{\textbf{(Color online) Graphene spintronic device and measurement
geometry. }(a) Schematic representation the graphene spintronic device
with ferromagnetic tunnel contacts on SiO$_{2}$/Si substrate for
a nonlocal (NL) and 3-terminal (3T) configuration. (b) Optical microscope
image of a multi-terminal spin-transport device showing a graphene
flake contacted by TiO$_{2}$ ($\unit[1]{nm}$)/Co/Au electrodes patterned
by electron beam lithography. }

\label{Device} 
\end{figure}

However, there has been a continuing discussion about spin parameters
obtained by the 3T technique. On the one hand, it has been reported
that the 3T geometry provides a lower limit for spin lifetimes and
larger magnitudes of the spin signal in semiconductors than theoretically
expected \cite{Dash2009,Dankert2013,Jansen2012a}. On the other hand,
spin lifetimes in metals obtained by 3T measurements are significantly
longer than theoretically predicted \cite{Txoperena2013}. Several
experiments also proposed an enhancement \cite{Tran2009,Jain2012b}
or inversion \cite{Dankert2013} of the 3T spin signal, whereas multiple
control experiments rule out any enhancement due to interface states
\cite{Jansen2012a,Sharma2014,Dash2009}. Therefore, it is required
to understand the contributing factors in both techniques by performing
experiments in the NL and 3T configuration using the same magnetic
tunnel contacts. Such measurements have been demonstrated for Si \cite{Sasaki2011,Suzuki2011,Sasaki2014},
GaAs \cite{Bruski2013} and epitaxial graphene on SiC substrate \cite{Birkner2013a}
at temperatures up to $\unit[100]{K}$, and on Ge \cite{Chang2013a}
up to room temperature. Even though there is a good agreement between
the NL and 3T spin parameters for most of those cases, some deviations
were observed for Ge at higher temperatures. This has been explained
by the fact that the 3T Hanle measurement is more easily affected
by additional scattering effects caused by the accompanied charge
current and the electric field under the FM contact. In spite of the
importance of studying those effects, especially at higher temperatures,
none of the other articles compares data above $\unit[100]{K}$. Although,
spin transport in exfoliated graphene has been studied the most, measurements
on different geometries has not been reported yet. 

Here we present spin transport and precession measurements in the
same exfoliated graphene device using both NL and 3T measurement geometries.
By analyzing the data on the basis of spin diffusion model we show
a very good agreement between the parameters obtained from both methods
up to room temperature. This indicates that there are no additional
enhancement or scattering mechanisms for the 3T geometry and validates
the applicability of both methods for spin transport and precession
measurements in graphene.

The graphene flakes were exfoliated from highly oriented pyrolytic
graphite (Advanced Ceramics), using the conventional cleavage technique,
onto a clean SiO$_{2}$ ($\unit[285]{nm}$)/highly doped n-type Si
substrate. The flake's thickness was characterized using a combination
of optical and atomic-force microscopy. In our experiments, we used
graphene flakes with a thickness of 2-3 layers, and a widths of around
$W=\unit[1.6]{\mu m}$. Ferromagnetic electrodes of different widths
($\unit[0.2-1]{\mu m}$) for the spin injector and detector were fabricated
by electron beam lithography, electron beam deposition and lift-off
technique. The shape anisotropy ensures different switching fields
of the electrodes allowing parallel and antiparallel configurations
through in-plane magnetic field sweeps. The injector and detector
electrodes were placed in a distance $L=\unit[2]{\mu m}$. The contacts
consist of $\unit[1]{nm}$ TiO$_{2}$ tunnel barrier, $\unit[65]{nm}$
Co and $\unit[20]{nm}$ Au capping layer. The TiO$_{2}$ barrier was
prepared by a twofold evaporation of $\unit[5]{\mathring{A}}$ of
Ti and oxidation in an oxygen atmosphere. This ensures a homogenous
and fully oxidized metaloxide barrier.

The electrical characterization of the device was performed using
a multi-terminal measurement geometry (Fig. \ref{Device}). The contact
resistance is about $R_{c}\approx\unit[2.3]{k\Omega}$ ($R_{c}A\approx\unit[600]{\Omega\mu m^{2}}$)
at $\unit[290]{K}$ with a nonmetallic temperature dependence indicating
a pinhole free tunnel barrier \cite{Jonsson-Akerman2000}. The channel
resistivity of the graphene was found to be $R_{\square}\approx\unit[1]{k\Omega}$.
Since $R_{c}>R_{\square}$ the back-flow of the injected spins into
the FM should be effectively suppressed \cite{Maassen2012b}. The
graphene channel showed a regular gate-dependent Dirac curve with
a mobility $\mu\approx\unit[2500]{cm^{2}\left(Vs\right)^{-1}}$. Spin
transport measurements were performed in both NL and 3T geometry using
a direct current (DC) of $\unit[5]{\mu A}$, whereas the voltage was
detected by a nanovoltmeter (Fig. \ref{Device}a). The measurements
were carried out under vacuum in a variable temperature cryostat with
a superconducting magnet.

\begin{figure}
\begin{centering}
\includegraphics[scale=1.25]{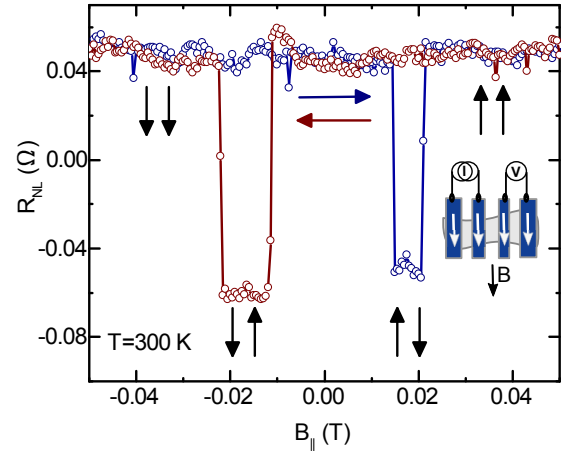}
\end{centering}

\protect\caption{\textbf{(Color online) Nonlocal spin-valve measurement.} Nonlocal
spin-valve signal measured at $\unit[300]{K}$ using an injection
current of $I=\unit[5]{\mu A}$. The sweep directions of the in-plane
magnetic field are indicated by blue (trace) and red (retrace). The
magnetic configurations of the electrodes are illustrated for both
sweep directions.}

\label{spinvalve} 
\end{figure}
The spin transport was studied in the nonlocal spin-valve geometry,
where the charge current path is isolated from the spin diffusion.
The spins injected through the Co/TiO$_{2}$ contacts accumulate in
the graphene, diffuse laterally and get detected by the nonlocal voltage
probes (Inset of Fig. \ref{spinvalve}). The nonlocal resistance is
recorded while the in-plane magnetic field is swept from a negative
to a positive value, followed by a reverse sweep. A nonlocal measurement
performed with a DC current $I=\unit[5]{\mu A}$ at room temperature
is shown in Fig. 2. A distinct spin-valve switching has been observed
between the parallel and antiparallel configurations of the injector
and detector electrodes, with a spin signal of $R_{NL}=\unit[110]{m\Omega}$.
This demonstrates the spin injection, transport, and detection
in our graphene device. 

\begin{figure}
\begin{centering}
\includegraphics[scale=1.25]{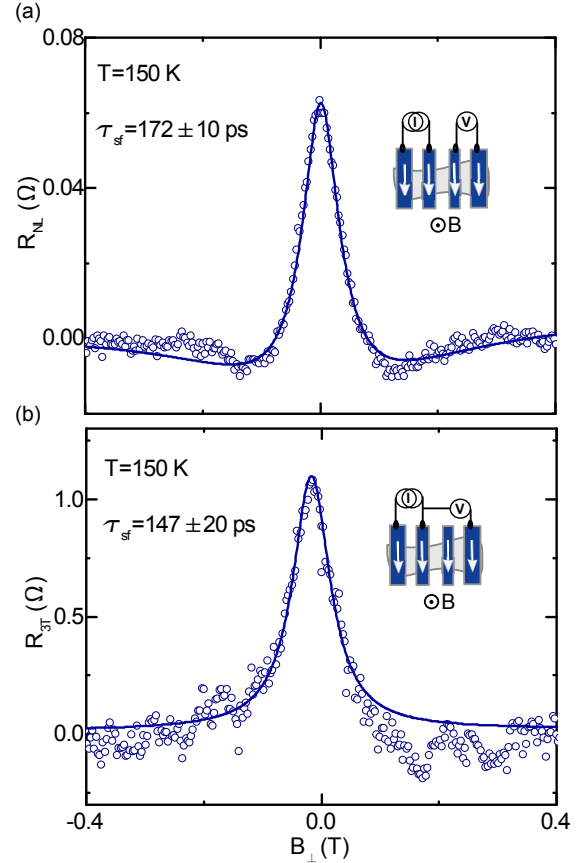} 
\end{centering}

\protect\caption{\textbf{(Color online) Hanle measurements in nonlocal (NL) and 3-terminal (3T) geometry.}
(a) NL Hanle spin-precession signal performed with an injection current
of $I=\unit[5]{\mu A}$ at $\unit[150]{K}$. The solid line represents
a nonlocal Hanle fit with spin drift, diffusion and precession equation
as in Eq. (\ref{eq:nonlHanle}) with a spin lifetime $\tau_{sf}=\unit[172\pm10]{ps}$
and $D=\unit[0.008]{m^{2}s^{-1}}$ . (b) 3T Hanle signal performed
with an injection current of $I=\unit[5]{\mu A}$ at $\unit[150]{K}$.
The solid line is the Lorenzian fit for 3T Hanle data as presented
in Eq. (\ref{eq:3Thanle}). }

\label{3T4Tcomp} 
\end{figure}

In order to evaluate the spin lifetime of the electrons in the graphene
device, we performed Hanle spin precession measurements in the NL
and 3T geometry. The NL Hanle spin precession measurement is performed
by sweeping the magnetic field perpendicular to the device geometry,
with the magnetization axes of the injector and detector electrodes
kept parallel (Inset Fig. \ref{3T4Tcomp}(a)). The injected spin-polarized
electrons precess around the perpendicular magnetic field $B_{\perp}$
with the Larmor frequency $\omega_{\text{L}}=g\mu_{\text{B}}B_{\bot}\hbar^{-1}$
(Lande's g-factor $g=2$) while diffusing towards the nonlocal detector
contact. The variation of this nonlocal resistance ($\Delta R_{NL}$)
due to precession and relaxation of the spins diffusing from the injector
to the detector can be described by Eq. (\ref{eq:nonlHanle}).\begin{widetext}
\begin{equation}
R_{NL}=\pm\frac{P^{2}R_{\square}}{W}\int_{0}^{\infty}\!\sqrt{\frac{D}{4\pi t}}\text{exp}\left[-\frac{L^{2}}{4Dt}\right]\text{cos}\left[\omega_{L}t\right]\text{exp}\left[-\frac{t}{\tau_{sf}}\right]\text{d}t\label{eq:nonlHanle}
\end{equation}
\end{widetext}With the measured sheet resistance $R_{\square}$
and predefined channel width $W$ and length $L$, we can extract
the spin polarization $P$ of the Co/TiO$_{2}$ contact, the diffusion
constant $D$, the spin lifetime $\tau_{sf}$.

Figure \ref{3T4Tcomp}(a) shows a nonlocal Hanle signal with a DC
injection current $I=\unit[5]{\mu A}$ at $T=\unit[150]{K}$. Fitting
the data with Eq. (\ref{eq:nonlHanle}) we obtain a spin lifetime
$\tau_{sf}=\unit[172\pm10]{ps}$ and a diffusion constant $D=\unit[0.008]{m^{2}s^{-1}}$
resulting in a spin diffusion length $\lambda_{sf}=\sqrt{D\tau_{sf}}=\unit[0.8]{\mu m}$
with a spin polarization $P=7\%$. The observed spin lifetime and
diffusion length are significantly shorter than theoretically expected
\cite{Ertler2009}, but compare well with the reported values for
graphene devices on SiO$_{2}$ using other oxide tunnel barriers \cite{Tombros2007}
with a similar carrier mobility of $\mu\approx\unit[2500]{cm^{2}\left(Vs\right)^{-1}}$.

Next we focus on Hanle measurements in the 3T configuration. The 3T
configuration is an extreme of a NL measurement scheme in which the
distance $L$ between injector and detector becomes zero. That means,
the same FM contact is used for spin injection and detection in the
graphene. The Hanle effect is used to control the reduction of the
induced spin accumulation by precession around an external perpendicular
magnetic field ($B_{\bot}$). The spin accumulation decays as a function
of $B_{\bot}$ with an approximately Lorentzian line shape given by

\begin{equation}
R_{3T}=\frac{P^{2}R_{\square}\lambda_{sf}}{2W\left[1+\left(\omega_{L}\tau_{sf}\right)^{2}\right]}.\label{eq:3Thanle}
\end{equation}
Figure \ref{3T4Tcomp}(b) shows a 3T electrical Hanle signal obtained
in the same Co/TiO$_{2}$/graphene contact as used for the NL measurement.
At a temperature of $\unit[150]{K}$ and a constant tunnel current
of $I=\unit[5]{\mu A}$ the spin resistance is found to be $R_{3T}=\unit[1]{\Omega}$,
with a polarization $P=8.1\%$ and a spin lifetime $\tau_{sf}=\unit[147\pm20]{ps}$.
This demonstrates that both NL and 3T measurements show a comparable
spin lifetime and polarization in our graphene device.

\begin{figure}
\begin{centering}
\includegraphics[scale=1.25]{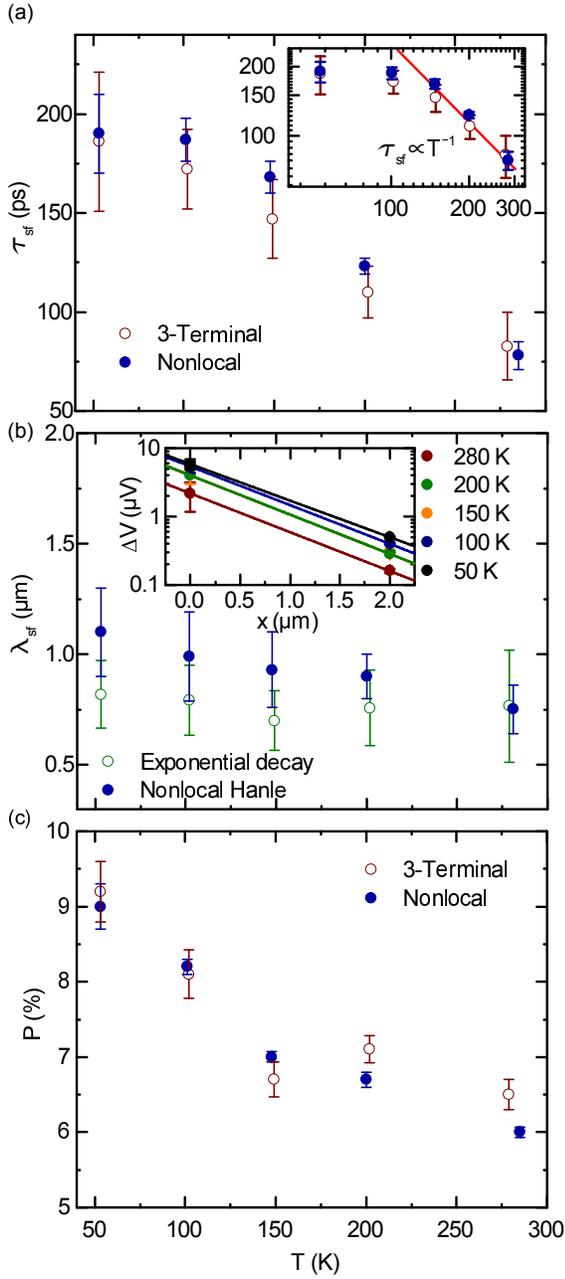} 
\end{centering}

\protect\caption{\textbf{(Color online) Comparison of spin parameters for nonlocal and 3-terminal Hanle.} (a) Spin lifetimes extracted directly from Hanle fits according
to Eq. (\ref{eq:nonlHanle}) and (\ref{eq:3Thanle}). Inset: Log-log
dependence of the main plot indicating a $\tau_{sf}\propto T^{\alpha}$
dependence (red line) with $\alpha=-1\pm0.1$ for $T\geq\unit[150]{K}$.
(b) Spin diffusion length as extracted from the nonlocal Hanle fit,
compared to the exponential decay of the spin signal amplitude (Inset)
from the injector to the detector electrode. (c) Spin polarization
extracted directly from Hanle fits according to Eq. (\ref{eq:nonlHanle})
and (\ref{eq:3Thanle}).}

\label{tau_lambda} 
\end{figure}
In order to understand the mechanisms affecting the spin accumulation
in both configurations we compare NL and 3T spin parameters at different
temperatures (Figure \ref{tau_lambda}). The spin lifetime is found
to be identical in the studied temperature range from $\unit[50-290]{K}$
(Fig. \ref{tau_lambda}(a)). Noticeably, the lifetime significantly
decreases for temperatures above $\unit[150]{K}$. This has been observed
previously in Si\cite{Sasaki2011} and Ge\cite{Chang2013a} and could
be attributed to phonon induced scattering. In the case of graphene,
such a distinct behavior has been previously reported by studying
the temperature dependence of the mobility in single and few layer
graphene devices \cite{Chen2008}. Above a temperature of $\unit[100-200]{K}$
the mobility was also strongly limited by impurities in the graphene
itself and significantly below the limit of longitudinal acoustic
and polar optical phonons of the SiO$_{2}$ substrate. Assuming $\tau_{sf}\propto T^{\alpha}$,
a power-law dependence of the spin lifetime on the temperature \cite{Chang2013a},
we extracted a coefficient $\alpha=-1\pm0.1$ for temperatures $T\geq\unit[150]{K}$.
In contrast to the results observed in Ge \cite{Chang2013a}, our
matching spin lifetimes, at low as well as for high temperatures, indicate that we have no additional scattering
mechanism in neither the 3T nor the NL technique. It has to be mentioned
that previous temperature dependent studies on single-layer (SLG)
and bi-layer (BLG) graphene reported significantly higher spin lifetimes
for low temperatures with different decays when warming up to room
temperature \cite{Han2011a}. The difference in temperature dependence
is proposed to stem from a drastic change in the scattering mechanism
going from SLG and BLG. Those studies used TiO$_{2}$ seeded MgO tunnel
barriers. In contrast, our values for the spin lifetime correspond
very well with other studies on regular metaloxide tunnel barriers
\cite{Tombros2007} and are similar to the temperature dependence
of BLG with a TiO$_{2}$/MgO tunnel barrier \cite{Han2011a}. However,
the scattering mechanism and spin lifetime can be also affected by
contact induced relaxation \cite{Maassen2012b}. 

 The amplitude of the spin signal, given as 

\begin{equation}
R(x)=\frac{P^{2}R_{\square}\lambda_{sf}}{2W}\text{exp}\left(-\frac{L}{\lambda_{sf}}\right),\label{eq:RNL_decay}
\end{equation}
depends exponentially on the distance $L$ between the injector and
detector electrode. If the measurement current and detected polarisation
are identical in both techniques, Eq. (\ref{eq:RNL_decay}) is simplified
to $V\left(x\right)=V\left(0\right)\cdot\text{exp}\left(-L\lambda_{sf}^{-1}\right)$.
Indeed we can show that the polarizations extracted from the 3T and
NL Hanle measurement are identical (Fig. \ref{tau_lambda}(c)). Employing
the simplified Eq. (\ref{eq:RNL_decay}) we extracted the spin diffusion
length $\lambda_{sf}$ (Inset Fig. \ref{tau_lambda}(b)), which is
identical to the value extracted from the NL fit with eq. (\ref{eq:nonlHanle}).
This confirms that the observed spin signal amplitude of the 3T measurement
correlates well to the signal of the NL method and rules out any form
of spin signal enhancement due to interface effects \cite{Dankert2013}.

In conclusion, we experimentally demonstrated the spin transport and
precession in graphene by studying the Hanle effect in NL and 3T measurement
geometries. We observe identical spin lifetimes, diffusion length
and polarization over a wide temperature range. The spin lifetime
decays from about $\unit[180]{ps}$ to $\unit[80]{ps}$ with a power-law
dependence for temperatures above $\unit[150]{K}$. The matching lifetimes from
both techniques rule out any additional scattering mechanisms for
either of the techniques. Furthermore, we were able to demonstrate
that the magnitude of the spin signal follows an exponential decay
with distance, as shown by the 3T and NL measurements. This rules
out any spin signal enhancement due to interface effects in the 3T
configuration. This verifies the applicability of both methods for
spin transport and precession measurements in graphene allowing for
faster developing and studying of future devices and geometries. Furthermore,
these results are of great significance for evaluating spin parameters
in other semiconducting two-dimensional materials \cite{Dankert2014}.

\textbf{Conflict of interests: }The authors declare no competing financial
interests.

\textbf{Acknowledgement:}The authors acknowledge the support of colleagues
at the Quantum Device Physics Laboratory and Nanofabrication Laboratory
at Chalmers University of Technology. The authors would also like
to acknowledge the financial supported from the Nano Area of the Advance
program at Chalmers University of Technology.

%


\end{document}